# Impact of Annealing on Perpendicular Magnetic Anisotropy in W/MgAl$_2$O$_4$/CoFeMnSi/W/CoFeMnSi/MgAl$_2$O$_4$/W Double Storage Layers for Upcoming MTJs

L. Saravanan[1*], Nanhe Kumar Gupta[2], Vireshwar Mishra[2], Sujeet Chaudhary[2], and Carlos Garcia[1,3]

[1]*Department of Physics, Technical University Federico Santa Maria, 2390123 Valparaiso, Chile*
[2]*Thin Film Laboratory, Department of Physics, Indian Institute of Technology Delhi, New Delhi 110016, India*
[3]*Centro Científico Tecnologico de Valparaíso - CCTVal, Universidad Tecnica Federico Santa María, 2390123 Valparaíso, Chile*

[*]E-mail: saravanan.lakshmanan@usm.cl; carlos.garcia@usm.cl

## Abstract

In this investigation, we successfully enhanced the uniaxial perpendicular magnetic anisotropy (PMA) in a W/MgAl$_2$O$_4$/CoFeMnSi/W/CoFeMnSi/MgAl$_2$O$_4$/W multilayer by adjusting the annealing temperature ($T_A$) [350ºC, 450ºC, and 550ºC]. At the specified $T_A$, we observed a maximum effective PMA energy density ($K_{eff}$) of ≈ 1.604 × 10$^6$ erg/cc accompanied by a low saturation magnetization ($M_s$). The enhancement of $K_{eff}$ with $M_s$ is significantly influenced by structural variations at the interfaces of CoFeMnSi and MgAl$_2$O$_4$, attributed to interfacial oxidation during adjustments in $T_A$. The $T_A$ was identified as a critical factor affecting the surface morphology, grain size, and surface roughness of the multilayer. Fourier-transform infrared (FTIR) measurements were employed to confirm the presence of Co-O or Fe-O vibrational bonds in the multilayer structures, elucidating the true origin of PMA. The control of interfacial oxidation at the interface during $T_A$ is crucial for regulating the strength of PMA. Therefore, this double CoFeMnSi/MgAl$_2$O$_4$-based multilayer presents a promising avenue, serving as a favourable candidate for future p-MTJs-based spintronic devices with enhanced thermal stability.

# I. INTRODUCTION

The development of non-volatile magnetic random-access memories (MRAM) is among the most promising advancements for spintronics applications. The core cell structure of this MRAM relies heavily on magnetic tunnel junctions (MTJs), which consist of a thin insulating layer serving as a tunnel barrier sandwiched between two ferromagnetic (FM) layers. The magnetic configuration of the FM layers dictates the electrical resistance and its intrinsic resistance is considered to encode the digital data stored in the MTJs. Over the years, successive generations of MRAM have undergone enhancements. Currently, the predominant generation is Spin-transfer torque MRAM (STT-MRAM) [1]. However, the STT-driven writing process demands a large critical switching current density [$10^6$-$10^7$ A/cm$^2$], leading to severe reliability and endurance issues due to tunnel barrier failures [2,3]. To address this challenge, spin-orbit torque (SOT) based MRAM has been introduced [4]. The spin-orbit torque (SOT) phenomenon can emerge from spin accumulation and uniform magnetization, especially in materials with strong spin-orbit coupling (SOC). An effective manipulation of the magnetization direction for the advancement of spintronic devices [5-9] has been demonstrated both theoretically and experimentally. In SOT-based MTJs, the FM (free/storage) layer directly interacts with the heavy metal (HM). During the writing process, a charge current is introduced into the HM layer, inducing a perpendicular spin current into the adjacent FM layer. This production of SOT results in the reversal of magnetization. The reading process relies on tunnelling magnetoresistance (TMR) via the vertical MTJ [10, 11]. The incorporation of different paths for reading and writing in MRAM leads to quicker access times and ultra-low power consumption. This makes SOT-MRAM cells particularly desirable for on-chip cache applications [12, 13]. Consequently, SOT-based MRAM stands out as one of the emerging technologies for the modern generation of memory devices [14].

One important crucial factor in p-MTJ devices is thermal stability (Δ), a parameter that influences data retention in the recording/storage/free layer. It can be expressed as $\Delta = K_{eff}V/k_BT$, where $K_{eff}$ is the effective Perpendicular Magnetic Anisotropy (PMA) energy, V is the volume of the storage layer, $k_B$ is the Boltzmann constant, and T is the temperature. To ensure the non-volatility of operational magnetic memories, a value of Δ greater than 60 is required [15]. To maintain robust thermal stability, a large PMA energy (high $K_{eff}$ value) becomes essential as the element size is reduced. Sufficient PMA energy can be achieved by utilizing bulk magnetic anisotropy, especially in $L1_0$-ordered (Co,Fe)-(Pt,Pd) alloys, (Co,Fe)/(Pt,Pd) multilayers, and rare-earth (RE)/transition metal (TM) multilayers. However, these materials may not be suitable as a storage layer in MTJs with oxide layers, particularly MgO, due to band mismatch, hindering the process of coherent spin-dependent tunneling. Realistically, interfacial PMA at the HM/FM/Oxide interface is employed to establish a perpendicular magnetic easy axis while maintaining a significant TMR effect [16]. In an HM/FM/Oxide structure, the PMA at the interface could be originated from band hybridization between the Co/Fe-$3d_{z^2}$ and O-$2p_z$ orbitals [17]. Therefore, the annealing temperature and film thickness need to be controlled sufficiently to overcome the demagnetizing energy of the storage layer. Enhancing the interfacial magnetic anisotropy is crucial for achieving a large Δ, not only for voltage-controlled MRAM [18, 19] but also for spin-torque-based MRAM devices [20]. Various approaches have been explored to enhance Δ using interfacial PMA, such as; introducing a different suitable buffer layer of HM [21, 22], a buffer layer with nitrogen doping [23], a spinel (MgAl$_2$O$_4$ - MAO) barrier [24], HM doping in an FM layer [25, 26], and an oxide (MgO) double barrier structure [27, 28].

To further enhance the Δ value, it is necessary to increase the thickness of the storage layer while maintaining robust PMA. Recent studies have explored the double interface storage structure of Oxide/FM/X/FM/Oxide, where X denotes the spacer or bridging layer. By adjusting the thickness of the interlayer between the two FM films, the coupling of the interlayer can be tuned to either antiferromagnetic or ferromagnetic states [29]. The introduction of a spacer layer between the two FM films has significantly improved the interfacial PMA property after annealing [30]. Notably, large PMA energy has been achieved in Oxide/FM/X/FM/Oxide-based structures, such as MgO/CFB/Ta/CFB/MgO [25], MgO/CFB/W/CFB/MgO [32-34], and MgO/CFB/Mo/CFB/MgO [35, 36]. Spacer/bridging layers of Ta, Mo, and W have been employed to enhance the PMA of the storage layer, particularly to meet the thermal budget of 400ºC. Among these spacer materials, tungsten (W) stands out due to its large Spin Hall Angle (θ$_{SHA}$) [37, 38], which enhances the effective PMA energy and imparts high thermal robustness to the storage layer, even up to 570ºC [39]. The insertion of a W film contributes to superior magnetic anisotropy with thermal stability [32].

On the other hand, the selection of FM and oxide layers is crucial for achieving strong PMA, a very low Gilbert damping constant (α), a Giant TMR ratio, and other desired properties. Thus far, double storage layer structures have been reported using CoFeB and MgO layers for FM and oxide layers, respectively. However, there are no reports on the utilization of Cobalt-based Heusler alloy (HA) and oxide layers compatible with PMA in Oxide/FM/X/FM/Oxide structures. A significant drawback of thin CoFeB films is the high α, which is inversely proportional to the CoFeB layer thickness [40]. To address this issue, a Co-based HA, particularly CoFeMnSi (CFMS), is proposed. Among various HA systems, the Equiatomic Quaternary Heusler alloys (EQHA) of CFMS have several advantages; high responsiveness to applied fields [41], spin-gapless semiconductor (SGS) behavior, structural stability surpassing the half-metallic ferromagnetic nature with very low power dissipation [43-45], strong PMA [46, 47], convenient substitutability for Diluted Magnetic Semiconductors (DMS), and potential suitability for voltage-controlled devices [48] and hydrogen sensors [42]. Recently, Zhang *et al.* reported PMA in multilayers of Ta/Pd/CFMS/MgO/Pd structures [42]. However, MTJs with MgO layers could suppress the TMR ratio due to improper lattice matching between the FM and MgO [49]. Therefore, there is a need to explore new oxide materials to match the FM properties of CFMS. One promising candidate is Metal Aluminate Oxide (MAO), known for its good lattice match with Cobalt-based HA compounds [50].

In this study, we explore a novel Oxide/FM/X/FM/Oxide structure composed of W, CFMS, and MAO, aiming to achieve interfacial uniaxial PMA with a low $M_s$ at elevated $T_A$. This endeavor is critical, as we discuss the role of the $T_A$ in manipulating the interfacial PMA, a key factor for enhancing the performance of advanced p-MTJs-based spintronic devices with robust thermal stability. Therefore, while PMA thin films with large $M_s$ values find application in bit-patterned media and perpendicular magnetic recording, our focus shifts towards achieving a strong PMA with a small $M_s$ value. This shift is deliberate, as it aligns with the requirements for low-power-consuming spin torque-based spintronic devices, where the strong PMA property with small $M_s$ values is preferred. In this research work, we manipulate the $T_A$ in p-MTJs, exploring the use of W-based multilayer material at a higher $T_A$ of 450ºC and showing the material's potential as a candidate for SOT-based novel spintronic devices. Consequently, the choice of materials and the strategic manipulation of $T_A$ become pivotal in our studies, aiming to offer a pathway to spintronic devices that are not only thermally stable but also energy efficient. The interplay between the $T_A$ and the magnetic properties of the Oxide/FM/X/FM/Oxide structure reveals intricate details about how to achieve the desired balance between PMA and $M_s$. This balance is crucial to impact the device's performance and energy consumption, making our research relevant for the next generation of spintronic technologies and providing valuable insights into material compositions and processing conditions that optimize magnetic properties for advanced applications.

# EXPERIMENT

All W(10)/MgAl$_2$O$_4$(1)/CoFeMnSi(2)/W(0.8)/CoFeMnSi(2)/MgAl$_2$O$_4$(1)/W(3) multilayer films were prepared on Si/SiO2 substrates at room temperature (RT) using a pulsed magnetron sputtering system *[Excel Instruments]*. The numbers in parentheses represent the nominal thickness of each layer in nanometers. The base pressure of the high vacuum sputtering chamber was maintained below $5 \times 10^{-7}$ Torr. W and CoFeMnSi (CFMS) were sputtered using direct current (DC) with powers of 40 W and 50 W, respectively. MgAl$_2$O$_4$ (MAO) was deposited by radio frequency (RF) sputtering with a power of 150 W. The sputtering rates for W, CFMS, and MAO were 0.40 Å/s, 0.69 Å/s, and 0.07 Å/s, respectively. All samples underwent *ex-situ* annealing at different temperatures ranging from 350°C to 550°C for 60 minutes without a magnetic field, where the base pressure of the annealing chamber was greater than $10^{-6}$ Torr.

The Grazing Incidence X-ray diffractometer (GI-XRD) with a Cu Kα source [λ= 1.5418 Å] [*PANalytical X'pert PRO*], X-ray reflectivity (XRR) [*PANalytical X'pert PRO and simulated by X'Pert Reflectivity, v1.2a* software], Atomic Force Microscopy (AFM) *[Asylum Research MFP3D-AFM]*, Superconducting Quantum Interference Device (SQUID) [*Quantum Design MPMS XL7*], and Fourier Transform Infrared spectrometer (FT-IR) [*Thermo Fisher Scientific Nicolet 6700*] were employed to characterize the crystal structure, confirm the thickness of each layer, analyze surface morphology, investigate magnetic properties, and identify the presence of metal and metal oxide bonds in all the multilayer films, respectively.

## II. RESULTS AND DISCUSSION

The schematic of W/MAO/CFMS/W/CFMS/MAO/W structures deposited on thermally oxidized silicon substrates is illustrated in Fig. 1(a). XRR curves and simulations for the as-deposited W(10)/MAO(1)/CFMS(2)/W(0.8)/CFMS(2)/MAO(1)/W(3) multilayer are presented in Fig. 1(b). Specular XRR data was used to determine film thickness by matching the simulated XRR profiles with the experimentally observed data.

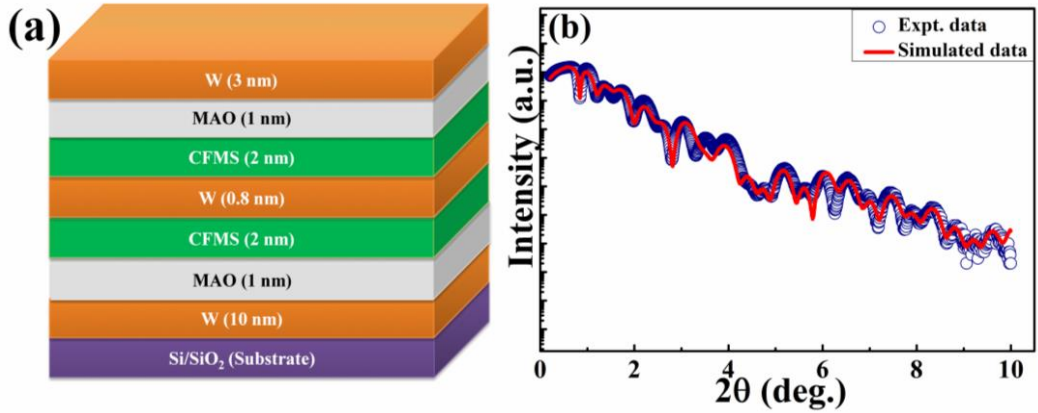

Fig. 1 (a) Schematic illustration of the W/MAO/CFMS/W/CFMS/MAO/W multilayer, (b) Reflectivity curves for the as-dep. W(10)/MAO(1)/CFMS(2)/W(0.8)/CFMS(2)/MAO(1)/W(3) sample.

To determine the crystalline structure of CFMS, MAO, and W films in our multilayer stacks, we conducted Grazing Incidence-X-ray Diffractometer (GI-XRD) studies on both as-deposited and annealed samples at 350°C, 450°C, and 550°C of Si/SiO$_2$//W/MAO/CFMS/W/CFMS/MAO/W at a glancing angle of 0.7°. The results are presented in Fig. 2. In the as-deposited film, the mian diffractions corresponding to the (110) plane of W were recorded, indicating partial crystallization of the W film. However, after annealing, additional diffractions, specifically (200), (211), and (220), were observed. In the XRD analysis, all peaks corresponding to thermodynamically stable body-centered cubic (bcc) α-W (space group Im-3m) phases were identified [51, 52]. Hence, it is reasonable to conclude that W films exhibit well-ordered cubic crystalline structures in W/MAO/CFMS/W/CFMS/MAO/W configurations when annealed above 350°C. However, changes in the crystallization of CFMS and MAO could not be discerned from the XRD patterns due to the ultra-low film thickness and the technical limitations of the XRD instrument. High-resolution transmission electron microscopy (HR-TEM) can provide deeper insights into the extent of crystallinity in these multilayers. Therefore, a more detailed exploration of the precise microscopic origin of such variation requires further experiments and analyses.

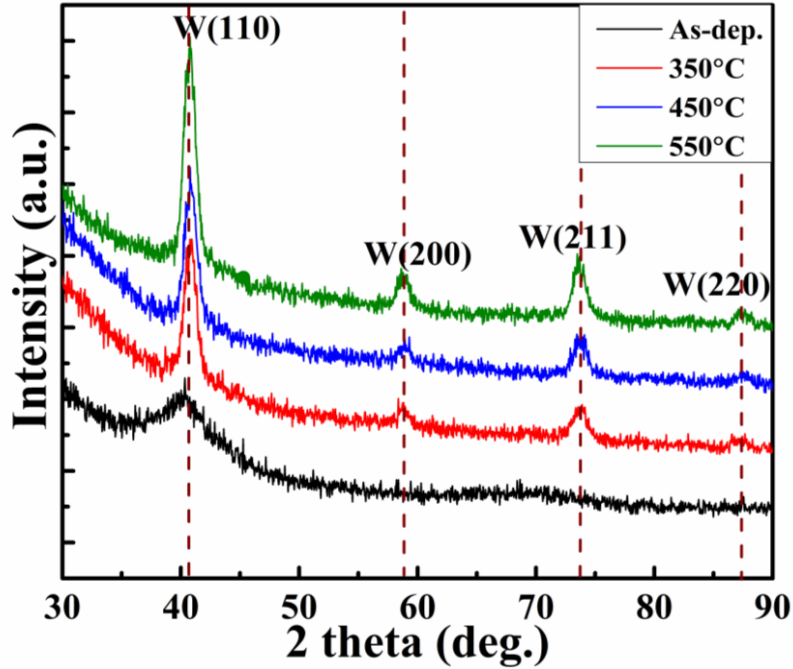

Fig. 2 GI-XRD patterns were obtained for both as-dep. and annealed W/MAO/CFMS/W/CFMS/MAO/W multilayer films across the temperature range up to 550ºC. The dashed lines in the patterns indicate the expected Bragg positions for bulk W.

1. Atomic Force Microscopy (AFM) analysis:

Fig. 3 shows the diverse surface morphologies of multilayers annealed at temperatures up to 550°C. The analysis covered a 1 × 1 μm² area (scale range: 200 nm) on the top surface of each sample. Surface morphology, surface roughness ($R_{rms}$), and particle size distributions of annealed multilayer samples were investigated using AFM in tapping mode. AFM images were captured at a resonance frequency ($f_r$) of 300 kHz, employing Al back-coated silicon cantilevers with a force constant of 40 N/m. The tip radius was less than 10 nm. In the as-dep. sample, see Fig 3.a, individual particles are scarcely visible, and there is no clear distinct separation between spherical-shaped particles. In contrast, annealed samples exhibited significantly different topographic images. At a temperature of 350°C, se Fig 3.b, particles grew densely, with a nearly uniform distribution and very few larger particles. For the 450°C sample, see Fig 3.c, we observed a homogeneous distribution of densely packed, spherical particles. Further increasing the $T_A$ to 550°C led to uneven sizes of neighboring particles as depicted in Fig. 3 (d). Particle sizes of ≈ 40 nm, ≈ 44 nm, and ≈ 47 nm were determined for the as-dep., 350°C, and 450°C samples, respectively. The 550°C sample exhibited a range of particle sizes, from ≈ 26 nm to ≈ 67 nm. Similarly, all samples displayed $R_{rms}$ values of ≈ 1.5008 nm, ≈ 0.2571 nm, ≈ 2.2507 nm, and ≈ 1.5728 nm. The particle size and $R_{rms}$ could significantly influence the magnetic anisotropy in the out-of-plane direction of the films [53, 54].

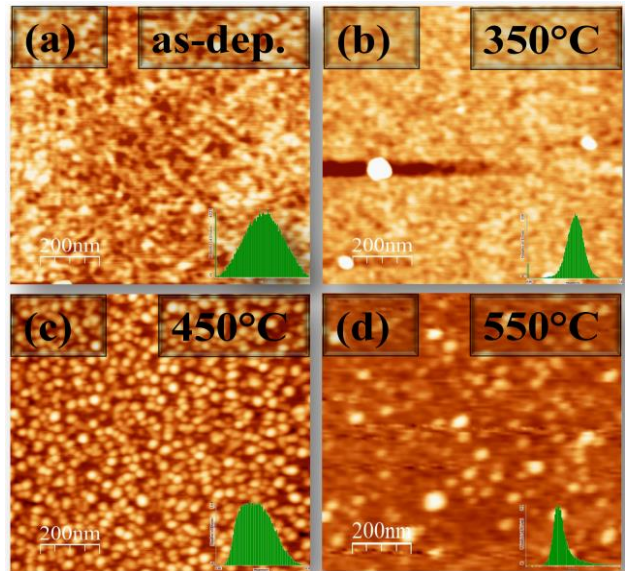

**Fig. 3** AFM images were captured for all double CFMS/MAO-based multilayer films at various temperatures. The inset figure shows the $R_{rms}$ histogram.

PMA is primarily influenced by the degree of interfacial oxidation at the heavy metal (HM)/ferromagnet (FM) and/or oxide/FM interfaces, due to the redistribution or migration of oxygen atoms [46, 47, 56, 57]. Therefore, it is crucial to optimize the annealing temperature.

The sample magnetic moment and its dependence on the magnetic field when applied parallel or perpendicular to the plane film was used to measure, with a resolution of $10^{-7}$ emu [55], by using a superconducting quantum interference device (SQUID). Figure 4 (a)-(d) shows the in-plane and out-of-plane M-H curves of W/MAO/CFMS/W/CFMS/MAO/W multilayer films annealed at different temperatures and measured at 300 K. The as-dep. sample exhibits in-plane magnetic anisotropy (IPA), see Fig. 4.a. After annealing the sample at 350ºC, the easy axis of magnetization persists in the parallel plane direction while a slight decrease of the anisotropy field ($H_k$) is observed. The presence of IPA in the multilayer films is attributed to the stronger demagnetization energy compared to interfacial perpendicular anisotropy, observed in both the as-dep. and 350ºC samples. However, interfacial PMA can be successfully achieved by tuning the annealing temperature. When the annealing temperature is further increased to 450ºC (Fig. 4(c)), the sample exhibits PMA along with IPA. The coexistence of both anisotropies arises from the interplay of interface and shape anisotropies in the film. Previous theoretical reports showed that the actual source of interfacial PMA at the FM/Oxide and/or Oxide/FM interface is related to the hybridization between the FM-3d$_z^2$ and O-2p$_z$ orbitals [17]. The sample annealed at 550ºC, depicted in Fig. 4(d), displays a rectangular hysteresis loop in the parallel plane direction indicating a strong resurgence of IPA, which can be attributed to an increase in demagnetization energy, surpassing the interfacial perpendicular anisotropy. It is noted that overoxidation at the interfaces during annealing can significantly degrade the PMA property [17], and numerous research works focus on creating multilayer structures with PMA properties [17, 46, 47, 58, 59].

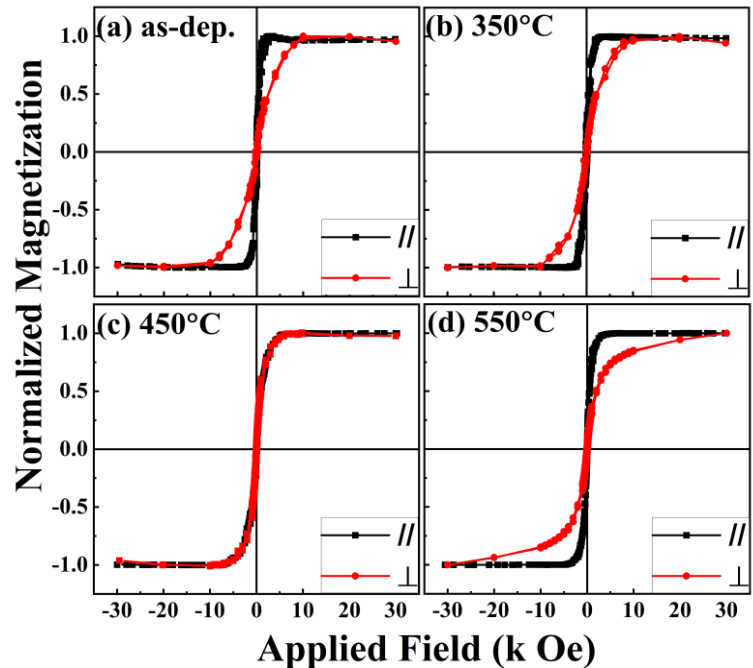

**Fig. 4** M-H loops were measured along both in-plane and out-of-plane directions for the W/MAO/CFMS(2)/W(0.8)/CFMS(2)/MAO/W multilayer at various annealing temperatures.

H$_k$ values were determined from the intersection of in-plane and out-of-plane saturated hysteresis loops [60, 61]. The positive and negative signs indicate the presence of PMA and IPA, respectively. The strength of H$_k$ is notably influenced by the T$_A$, the obtained H$_k$ values are -9.99 kOe, -9.44 kOe, 6.24 kOe, and -27.44 kOe for the as-dep., 350ºC, 450ºC, and 550ºC samples, respectively (Fig. 5(a)). Uniaxial PMA is exclusively achieved in the multilayer films at T$_A$ of 450ºC. Generally, the presence of Bloch-type domain walls (with skinny wall thickness) and Neel-type domain walls (with a large wall thickness) favors PMA and IPA properties, respectively [62]. In our study of multilayer annealed at 450ºC, the emergence of uniaxial PMA could stem from a combination of Bloch and Neel-type domain walls. When the T$_A$ is adjusted from > 350ºC to ≤ 450ºC, it is plausible that Bloch-type domain walls are more significantly formed, facilitated by an adequately improved oxidation state with a strong interfacial PMA.

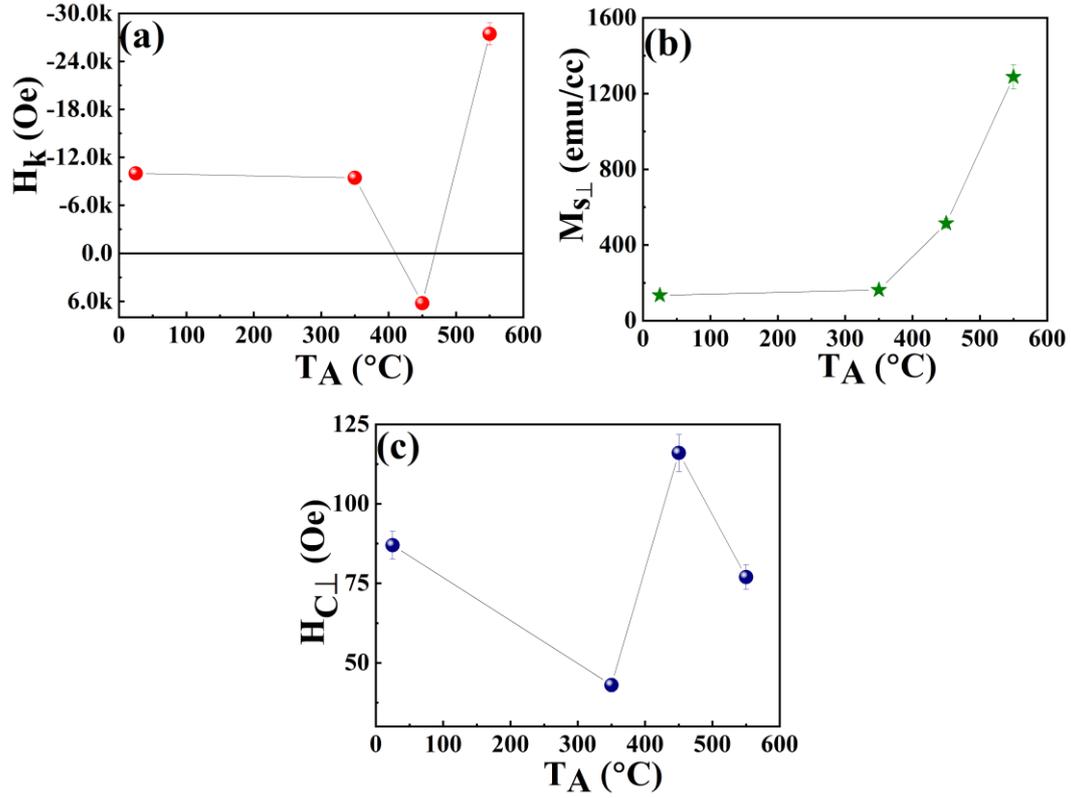

**Fig. 5.** The $T_A$ dependence of (a) $H_k$, (b) $M_{s\perp}$, and (c) $H_{c\perp}$ is investigated for the double CFMS/MAO-type multilayer.

The out-of-plane saturation magnetization ($M_{s\perp}$) as a function of $T_A$ is shown in Fig. 5(b). The in-plane $M_s$ values steadily increase from ≈ 121 emu/cc to ≈ 3047 emu/cc with an elevation in $T_A$. A similar trend is observed for the double CFMS/MAO layers-based multilayer films, where the out-of-plane Ms values gradually rise from ≈ 134 emu/cc to ≈ 1288 emu/cc for the multilayer films annealed at 350ºC, 450ºC, and 550ºC, respectively. The increase in $M_s$ values in the multilayer structure could originate from the development of the crystalline structure, transitioning from the disordered phase (A2) to the partially ordered (B2) or fully ordered phases (L2$_1$) in the Heusler alloys [63]. The oxide layer could significantly impact the Ms values at the CFMS (2 nm)/MAO(1 nm) and/or MAO(1 nm)/CFMS(1 nm) interfaces. Therefore, the higher Ms values are attributed to the presence of an optimal interfacial oxidation state and low bulk anisotropy, resulting in a coherent lattice match at the MAO/CFMS and CFMS/MAO interfaces. Consequently, under-oxidation/over-oxidation of the CFMS layers near the interfaces also plays a vital role in increasing in-plane and out-of-plane $M_s$ values. The observed enhanced interfacial oxidation state at the interfaces of the multilayer films is in good agreement with previous experimental reports [46, 47, 58].

Nevertheless, the observation of $M_s$ in the out-of-plane direction is notably lower than the previously recorded values for other cobalt-type PMA layers, such as Pt/Co ($\approx$ 900 emu/cc) [64], CoFeB films ($\approx$ 1200 emu/cc) [65], CoFe ($\approx$ 1200 emu/cc) [66] and Co/Ni ($\approx$ 660 emu/cc) [67]. These relatively small $M_s$ values are crucially required for ultra-low power-consuming spin torque type p-MTJs, as the intrinsic threshold current density ($I_{C0}$) is primarily dependent on $M_s$. The $I_{C0}$ is expressed as [54],

$$I_{C0} = \left[e\alpha \frac{M_S^2 \Delta}{g\mu_B P}\right] \gamma_0 \left[\frac{1}{|\beta - \alpha|}\right] \quad (1)$$

Here, $M_s$, $\Delta$, $\alpha$, $\beta$, $g$, $\mu_B$, $P$ and $e$ are the saturation magnetization of the magnetic film, domain wall thickness, Gilbert damping constant, non-adiabatic spin-transfer parameter, gyromagnetic ratio, Bohr magneton, spin polarization of the FM layers and charge of the electron respectively.

Figure 5(c) displays $H_{c\perp}$ for all double CFMS/MAO-based multilayer films at various $T_A$. $H_{c\perp}$ initially decreases when the $T_A$ of 350°C and then it rises rapidly at 450°C, reaching a maximum value of $\approx$ 116 Oe. Further increasing of the $T_A$ to 550°C slowly decreases the $H_{c\perp}$ ($\approx$ 77 Oe) of the multilayer. The variation in $H_{c\perp}$ could be attributed to the modifications in the magnetic pinning sites and the effects of over-oxidation at their interfaces [58, 70].

To investigate the strength of magnetic anisotropy in multilayer structures, we determined the effective perpendicular magnetic anisotropy energy density ($K_{eff}$). Negative and positive signs of $K_{eff}$ represent the presence of IPA and PMA, respectively [71]. $K_{eff}$ is given by,

$$K_{eff} = \frac{M_S \times H_K}{2} \quad (2)$$

Regarding magnetization, two essential factors are assumed for the PMA property of the films. One factor is volume magnetization, contributed by the volume magnetic moment with negative values of $K_{eff}$. The magnetization direction is oriented in the parallel plane configuration. Another factor is interface magnetization, originating from the interfacial magnetic moment with positive values of $K_{eff}$. The magnetization orientation is in the out-of-plane or perpendicular direction of the films. The correlation between the volume magnetic anisotropy ($K_v$) and interfacial magnetic anisotropy ($K_i$) constants is expressed as,

$$K_{eff} = K_V - 2\pi M_S^2 + \frac{K_i}{t_{CoFeMnSi}} \quad (3)$$

Equation (3) can be rewritten as,

$$K_{eff} \times t_{CoFeMnSi} = [K_V - 2\pi M_S^2] t_{CoFeMnSi} + K_i \qquad (4)$$

where $t_{CoFeMnSi}$ is the thickness of CoFeMnSi film and $2\pi M_s^2$ is the demagnetizing energy correlated with the thin film's shape anisotropy. Notably, the $K_v$ is not considered in the PMA property of the soft ferromagnetic (CoFeMnSi) multilayer structures. In this context, the $K_{eff}$ originates solely from interfacial anisotropy, resulting in a positive value. When the demagnetization energy is significant, the multilayer structure could exhibit IPA ($K_{eff} < 0$). Conversely, the multilayer films demonstrate PMA when interfacial out-of-plane anisotropy surpasses the demagnetization energy ($K_{eff} > 0$). The evaluated out-of-plane $M_s$, $H_c$, $H_k$, and $K_{eff}$ values of all the as-dep. and annealed samples are given in Table 1.

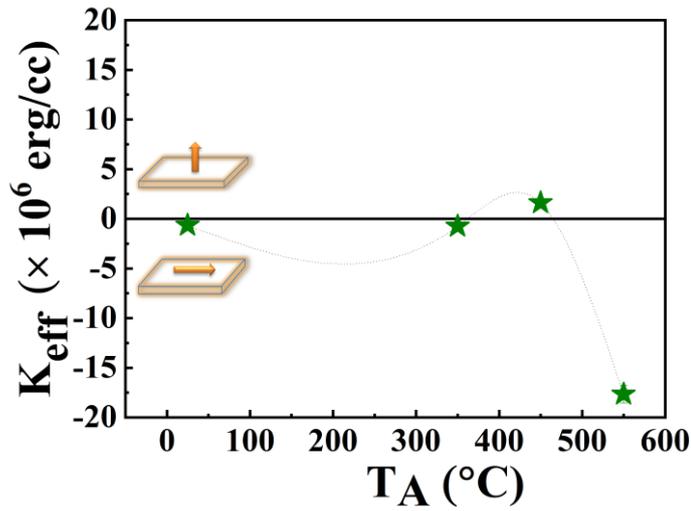

**Fig. 6.** $K_{eff}$ as a function of $T_A$ for the double CFMS/MAO-based multilayer films.

**Table 1:** Perpendicular magnetic properties of W/MAO/CFMS(2)/W(0.8)/CFMS(2)/MAO/W multilayer with various post annealing temperatures.

| S.No | Annealing temperatures (°C) | Magnetic properties | | | |
|---|---|---|---|---|---|
| | | $M_{s\perp}$ ($\approx$ emu/cc) | $H_{c\perp}$ ($\approx$ Oe) | $H_k$ ($\approx$ k Oe) | $K_{eff}$ ($\approx \times 10^6$ erg/cc) |
| 1 | As-dep. | 134 | 87 | -9.998 | -0.669 |
| 2 | 350 | 162 | 43 | -9.448 | -0.765 |
| 3 | 450 | 514 | 116 | 6.244 | 1.604 |
| 4 | 550 | 1288 | 77 | -27.445 | -17.67 |

Fig. 6 displays the variation of $K_{eff}$ with $T_A$. $K_{eff}$ increases from a negative value of $-0.669 \times 10^6$ erg/cc (IPA) to a maximum positive value of $1.604 \times 10^6$ erg/cc (PMA). Subsequently, it abruptly decreases to $-17.67 \times 10^6$ erg/cc (IPA) with increasing $T_A$. The observed maximum

$K_{eff}$ value in this study surpasses those reported for similar multilayer films annealed at 300ºC, such as Ta/Pd/CoFeMnSi (CFMS) (2.3 nm)/MgO (1.3 nm)/Pd (with $K_{eff} \approx 0.56 \times 10^6$ erg/cc) and Ta/Pd/Mn$_2$CoAl (1.7 nm)/MgO (with $K_{eff} \approx 0.2 \times 10^6$ erg/cc) multilayer films [42, 72]. Recently, we reported a significant $K_{eff}$ value of $\approx 0.35 \times 10^6$ erg/cc for the film MgAl$_2$O$_4$ (1.0 nm)/CoFeMnSi (2.0 nm)/MgAl$_2$O$_4$ (1.0 nm)/Ti (3.0 nm) at a higher $T_A$ of 400ºC [46]. Furthermore, the observed PMA energy of $1.604 \times 10^6$ erg/cc is slightly lower than the maximum PMA energy determined in the MgAl$_2$O$_4$ (1.5 nm)/ CoFeMnSi (2.0 nm)/ MgAl$_2$O$_4$ (1.0 nm)/W (3.0 nm) system at 400ºC [47].

According to theoretical reports, the real source of PMA comes from the orbitals of Co/Fe-3d and O-2p hybridization at Oxide/FM and/or FM/Oxide interfaces [73, 74]. As a result, oxidation in the multilayer films plays a significant role in influencing the PMA property. Fourier Transform Infrared Spectroscopy (FTIR) is a powerful tool for identifying the presence of metal and metal oxide vibration bands in samples. FTIR measurements for both the as-dep. and annealed structures are presented in Fig. 7.

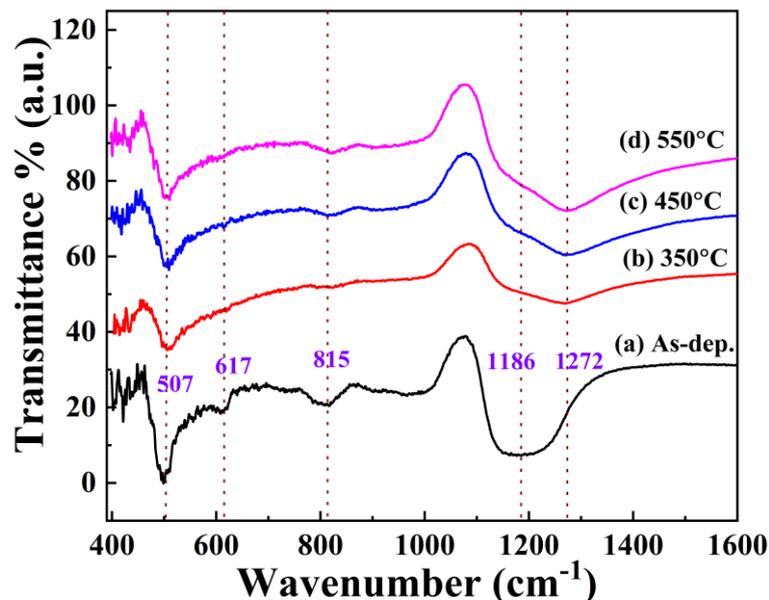

**Fig. 7** FTIR spectra of W/MAO/CFMS/W/CFMS/MAO/W multilayer samples at different conditions: (a) As-dep., (b) 350ºC, (c) 450ºC and (d) 550ºC.

The infrared spectrum was recorded for the double CFMS/MAO-based multilayer films in the range of 400-4000 cm$^{-1}$, with a focus on the fingerprint region. Analysis of this region aimed to identify the purity and presence of metal and metal oxide in the stacks. From the observed FTIR data, two primary vibrational bands were consistently detected at 507 cm$^{-1}$ and 815 cm$^{-1}$ for both as-dep. and annealed samples. The band at 507 cm$^{-1}$ is associated with the presence of randomly oriented octahedral, cobalt in an oxygen octahedral environment, specifically attributed to stretching vibration ($\nu_{Co-O}$) [75]. The band at 815 cm$^{-1}$ is assigned to the stretching mode of $\nu_{W-O}$ vibration, indicating the hexagonal structure of tungsten oxide (WO). A low wavenumber band at 617 cm$^{-1}$ was observed exclusively in the as-dep. sample and is attributed to the Al-O bond (α-Al$_2$O$_3$) of a pseudo-boehmite structure [76]. The presence of Al-O in the multilayer is suggested to have originated from the nearby layer of MgAl$_2$O$_4$. This suggests that Al may have been slightly separated from the MgAl$_2$O$_4$ layer during deposition. Interestingly, the band of Fe-O was not observed in both as-dep. and annealed samples, implying that the Co-O band vibration strongly dominated in all the double CFMS/MAO-based multilayer films. The as-dep. sample shows a coexistence of Co-O and Al-O bonding. At T$_A$ (450°C), only the contribution of the Co-O bond is observed and it is distinct from the as-deposited sample. The FTIR results suggest that the oxidation of Co at the CFMS/MAO and/or MAO/CFMS interface is the actual source of PMA.

On the contrary, excessive oxidation of Co could compromise the interfacial PMA [77, 78] at a higher T$_A$ of 550°C. This is attributed to the reduction in the number of mixed states (spin-orbit induced mixed states between the $\Delta_1$ majority and $\Delta_2$ minority Bloch states) involving both Co-3d$_Z^2$ and O-2p$_Z$ orbitals, crucial for PMA at the interfaces of MAO and CFMS. This reduction occurs due to the redistribution of local charge induced by surplus oxygen atoms [73]. In conclusion, based on FTIR measurements, we infer that W/MAO(1 nm)/CFMS(2 nm)/W(0.8 nm)/CFMS(2 nm)/MAO(1 nm)/W stacks at 450°C elucidate uniaxial PMA arising solely from the stretching mode of the vibrational bonding of Co-O (with suitable oxygen content) and not from the Fe-O vibrational band. Additionally, vibrational bands at 1186 cm$^{-1}$ for the as-dep. sample and 1272 cm$^{-1}$ for the annealed sample are observed and correspond to the IR spectra of SiO$_2$, specifically the stretching mode of $\nu_{Si-O}$, formed in the range of 1000 cm$^{-1}$ – 1500 cm$^{-1}$. The presence of compressive stress at the substrate Si/SiO$_2$ interfaces is identified as responsible for the variation in the IR spectra [79] between the as-dep. and annealed samples.

## III. SUMMARY AND CONCLUSION

In summary, we conducted a systematic investigation of the structural, topographical, magnetic, and metal oxide presence in W/MgAl$_2$O$_4$/CoFeMnSi/W/CoFeMnSi/MgAl$_2$O$_4$/W multilayer films grown on SiO$_2$ substrates by varying the T$_A$ up to 550°C. The particle sizes increased from ≈ 40 nm to ≈ 67 nm as T$_A$ rose from RT to 550°C. The interfacial PMA was notably responsive to $T_A$. A substantial PMA energy ($K_{eff}$) of ≈ 1.604 × $10^6$ erg/cc with a small $M_{s\perp}$ of ≈ 514 emu/cc was observed only at a higher $T_A$ of 450ºC, indicating the optimal level of interfacial oxidation. This points to preferential orbital hybridization at the interfaces of $MgAl_2O_4$/CoFeMnSi/W and/or W/CoFeMnSi/$MgAl_2O_4$. FTIR analysis confirmed the presence of solely Co-O (stretching vibrational bond) in the sample. Therefore, the uniaxial PMA in the W/$MgAl_2O_4$/CoFeMnSi/W/CoFeMnSi/$MgAl_2O_4$/W multilayer at $T_A$ of 450ºC is attributed to the hybridization of Co-$3d_{z^2}$ and O-$2p_z$ orbitals at their interface. Conversely, over-oxidation at their interfaces could destroy the PMA and lead to the development of IPA property, enhancing the $M_s$ values at 550ºC. Hence, the combination of robust PMA energy with a low $M_s$ value in the W(10 nm)/$MgAl_2O_4$(1 nm)/CoFeMnSi (2 nm)/W(0.8 nm)/CoFeMnSi(2 nm)/$MgAl_2O_4$(1 nm)/W(3 nm) multilayer at high thermal stability of 450ºC holds potential for the development of modern spintronic devices.

## ACKNOWLEDGEMENTS

L.S. acknowledges to FONDECYT Postdoctorado 2022 ANID, 3220373. C. Garcia acknowledges the financial support received by ANID FONDECYT/Regular 1201102, ANID FONDEQUIP EQM140161, ANID FONDEQUIP EQM 150094. This work was also supported by the European Union's Horizon 2020 research and innovation program under the Marie Sklodowska-Curie Grant: H2020-MSCA-RISE, ID: 734801 EU - DOI: 10.3030/734801 and H2020-MSCA-RISE, ID: 101007825 EU - DOI: 10.3030/101007825.

**Highlights:**

- Uniaxial PMA is achieved in the Si/SiO$_2$//W/MgAl$_2$O$_4$/CoFeMnSi/W/CoFeMnSi/MgAl$_2$O$_4$/W stacks at different annealing temperatures (T$_A$).

- Robust PMA energy (K$_{eff}$ ≈ $1.604 \times 10^6$ erg/cc) is observed with a low M$_{s\perp}$ of ≈ 514 emu/cc.

- Enhanced thermal stability with PMA is maintained at 450ºC.

- PMA is attributed to only the Co-O stretching mode of vibration, as supported by FTIR studies.